\documentclass[journal, 10pt, a4paper]{IEEEtran}
\IEEEoverridecommandlockouts
\ifCLASSINFOpdf
\usepackage[pdftex]{graphicx}
\else
\fi
\usepackage{amsmath}
\usepackage{array}
\ifCLASSOPTIONcompsoc
\usepackage[caption=false,font=normalsize,labelfont=sf,textfont=sf]{subfig}
\else
\usepackage[caption=false,font=footnotesize]{subfig}
\fi
\usepackage{fixltx2e}
\usepackage{stfloats}
\usepackage{cite}
\usepackage{url}
\usepackage{hyperref} 
\hypersetup{colorlinks = true, linkcolor=blue, citecolor=red, urlcolor=blue} 

\usepackage{amssymb}
\usepackage{amsthm}
\usepackage{algpseudocode}
\usepackage{algorithm}

\usepackage{graphicx}
\usepackage{mathptmx}
\usepackage{amsmath}
\usepackage{xcolor}
\allowdisplaybreaks

\usepackage{cuted}
\usepackage{multicol}
\DeclareMathOperator{\sinc}{sinc}
\usepackage{setspace}

\begin{document}
\title{Artificial Intelligence Assisted Collaborative Edge Caching in Small Cell Networks}

\author{\IEEEauthorblockN{Md Ferdous Pervej\IEEEauthorrefmark{1},  Le Thanh Tan\IEEEauthorrefmark{2}, and Rose Qingyang Hu\IEEEauthorrefmark{3}}\\
	\IEEEauthorblockA{\IEEEauthorrefmark{1}Department of Electrical and Computer Engineering, North Carolina State University, Raleigh, NC 27695, USA} \\
	\IEEEauthorblockA{\IEEEauthorrefmark{2}Commonwealth Cyber Initiative, Old Dominion University, Norfolk, VA 23529, USA} \\
	\IEEEauthorblockA{\IEEEauthorrefmark{3}Department of Electrical and Computer Engineering, Utah State University, Logan, UT 84322, USA} \\
	Email: \tt mpervej@ncsu.edu, tle@odu.edu, rose.hu@usu.edu 

\thanks{The work of M. F. Pervej, L. T. Tan and R. Q. Hu were supported in part by National Science Foundation under grants NeTS 1423348 and EARS 1547312 as well as in part by the Intel Corporation.}
\thanks{This is the \textbf{technical report} of \cite{Pervej_CHR}.}}

\maketitle

\begin{abstract}
Edge caching is a new paradigm that has been exploited over the past several years to reduce the load for the core network and to enhance the content delivery performance. 
Many existing caching solutions only consider homogeneous caching placement due to the immense complexity associated with the heterogeneous caching models. 
Unlike these legacy modeling paradigms, this paper considers heterogeneous content preference of the users with heterogeneous caching models at the edge nodes. 
Besides, aiming to maximize the cache hit ratio (CHR) in a two-tier heterogeneous network, we let the edge nodes collaborate.
However, due to complex combinatorial decision variables, the formulated problem is hard to solve in the polynomial time.
Moreover, there does not even exist a ready-to-use tool or software to solve the problem. 
We propose a modified particle swarm optimization (M-PSO) algorithm that efficiently solves the complex constraint problem in a reasonable time. 
Using numerical analysis and simulation, we validate that the proposed algorithm significantly enhances the CHR performance when comparing to that of the existing baseline caching schemes.
\end{abstract}


\IEEEpeerreviewmaketitle
	
\section{Introduction}
Owing to the ever growing requirements of enhanced data rates, quality of service, and latency, wireless communication has evolved from generation to generation.
With the exponential increase of the connected devices, existing wireless networks have already been experiencing performance bottleneck.  
While the general trends are shifting resources towards the edge of the network \cite{pervej2020dynamic, pervej2020eco, pervej_upledge}, study shows that mobile video traffic is one of the dominant applications that prompt this bottleneck \cite{6787081, molisch2014caching, siris2015multi}. 
Caching has become a promising technology to address  this performance issue by storing popular contents close to the end users \cite{8714022,8632730}. 
Therefore, during the peak traffic hours, the requested contents can be delivered from these local nodes ensuring a deflated pressure to the backhaul and the centralized core network yielding reduced latency for content delivery.
Thus, an edge cache-enabled network utilizes the much-needed wireless spectrum and wireline bandwidth efficiently. 
In the ultra-dense network platform, caching at the edge nodes is a powerful mechanism for delivering video traffic.

While the caching solution can significantly benefit next-generation wireless communication, it still comes with various challenges \cite{xu2004caching, sheng2016enhancement, wang2014cache, liu2016caching}. 
First of all, the content selection has an enormous impact on the cache-enabled platform \cite{8998261,pervej_upledge}. 
Then, choosing the appropriate nodes to store the contents needs to be answered. 
Due to the broad combinatorial decision parameters, this is an immense challenge for any cache-enabled network platform. 
Furthermore, owing to the necessity of the system performance metrics, the solution to this combinatorial decision problem may change. 
Therefore, based on the performance metric, an efficient solution is demanded to handle the issues in a reasonable time. 
As such, under practical modeling with proper communication protocols, a heterogeneous network platform needs to be adopted for evaluating the caching performance.

There exist several caching solutions in the literature \cite{8458381,8998261,tan2018d2d,8667875}. 
Caching policy and cooperative distance were designed in \cite{8458381}, by Lee \textit{et al.}, considering clustered device-to-device (D2D) networks. 
While the authors showed some brilliant concepts for the caching policy design aiming to maximize (a) energy efficiency and (b) throughput, they only considered the collaboration among the D2D users. 
Lee \textit{et al.} also proposed a base station (BS) assisted D2D caching network in \cite{8998261} that maximizes the time-average service rate. 
However, the authors only considered a single BS underlay D2D communication with homogeneous request probability modeling. 
Tan \textit{et al.} \cite{tan2018d2d} adopted the collaboration based caching model in the heterogeneous network model. 
A mobility aware probabilistic edge caching approach was explored in \cite{8667875}. 
The authors' proposed model considered the novel idea of collaboration by considering the spatial node distribution and user-mobility.
While \cite{tan2018d2d,8667875} introduces some splendid concepts of relaying and collaborations, the authors only incorporated homogeneous caching placement strategies.

Unlike these existing works, we investigate heterogeneous content preference model leveraging heterogeneous cache placement strategy in this paper.
Particularly, in a small cell network (SCN), we incorporate collaborations among spatially distributed full-duplex (FD) enabled BSs and half-duplex (HD) operated D2D users to maximize the average cache hit ratio (CHR).
However, the formulated problem contains intricate combinatorial decision variables that are hard to determine in polynomial time. 
Therefore, we implement a modified particle swarm optimization (M-PSO) algorithm that effectively solves the grand probabilistic cache placement problem within a reasonable time. 
To the best of our knowledge, this is the first work to consider heterogeneous user preference with a heterogeneous caching model in a practical SCN that uses collaborative content sharing among heterogeneous edge nodes to maximize the CHR.

The outline of this paper is as follows. 
The system model and the proposed content access protocols are presented in Section~\ref{System_Model_Problem_Defi}, followed by the  CHR analysis in Section~\ref{FD_EDGE_Cost_Analysis}.
The optimization problem and the  proposed M-PSO algorithm are described in Section~\ref{Opt_Prop_Algo}. 
Section~\ref{Result} gives the performance results, followed by the concluding remarks in Section~\ref{conclu}.

\section{System Model and Content Access Protocols}
\label{System_Model_Problem_Defi}
This section presents the node distributions and describes the caching properties,  followed by the proposed content access protocols.

\subsection{Node Distributions}

We consider a practical two-tier heterogeneous network, which consists of macro base stations (MBS) and low-power sBSs (or relays) with underlaid D2D users. 
The nodes are distributed following an independent homogeneous Poisson point processes (HPPP) model. 
Let us denote the densities of the D2D user, sBS and MBS by $\lambda_u$, $\lambda_b$ and $\lambda_m$, respectively.
The  sBSs and MBSs operate in the FD mode whereas the D2D users operate in the  HD mode.
Let us denote the set of D2D users, sBSs and MBSs by $\mathcal{U}$, $ \mathcal{B}$ and $\mathcal{M} $, respectively. 
Without any loss of generality, user, sBS and MBS are denoted by $u \in \mathcal{U}$, $b \in \mathcal{B}$, and $m \in \mathcal{M}$, respectively. 
Besides, the communication ranges of these nodes are denoted by $R_u$, $R_b$ and $R_m$, respectively.

The requesting user node is named as the tagged user node. 
While a user is always associated with the serving MBS, it can also associate with a low powered sBS if the association rules are satisfied.  The main benefits of being connected to sBS over MBS are higher data rate, less latency, less power consumption, more effective uses of radio resources, etc.
We denote the associated sBS as the tagged sBS for that user. 
Furthermore, if such a tagged sBS exists for the user, the user maintains its communication with the serving MBS via the tagged sBS. 
In that case, the sBS can also use its FD mode to deliver requested content from the other sBSs or the cloud via the MBS. 
If such a tagged sBS does not exist for the user, the user will have to rely on the neighbor sBS nodes and the serving MBS for extracting the requested contents. 
As all the users may not place a content request at the same time, we assume that only $\alpha$ portions of the users act as tagged users.
Without any loss of generality, the requesting user, the associated sBS, and the serving MBS are denoted as $u_0$, $b_0$ and $m_0$, respectively.

\subsection{Cache Storage, Caching Policy and Content Popularity}

The cache storage of the users, sBSs and MBSs are denoted by $\mathcal{C}_{d}, \mathcal{C}_b$ and $\mathcal{C}_m$, respectively. 
Considering equal-sized contents we investigate a probabilistic caching placement \cite{7248843} where the users can make a content request from a content directory of $\mathcal{F} = \{f_k\}$, where $k \in \{1,2,\dots,F\}$. 
For the caching model, a probabilistic method is considered assuming a heterogeneous caching placement strategy. 
Let $\eta_{f_k}^{u_i}$, $\eta_{f_k}^{b_j}$ and $\eta_{f_k}^{m_l}$ be the probabilities of storing a content $f_k \in\mathcal{F}$ at the cache store of the user node $u_i$, the sBS $b_j$ and the MBS $m_l$, respectively. 
Note that probabilistic caching is highly practical and adopted in many existing works \cite{6787081, molisch2014caching,8458381,8998261,tan2018d2d,8667875, 7248843}.

The content popularity is modeled by following the $\mathrm{Zipf}$ distribution with the probability mass function $\mathrm{P}_{f_k} = \frac{k^{-\gamma}}{\sum_{k=1}^{F} k^{-\gamma}}$.
Note that the skewness $\gamma$ governs this distribution. 
It is assumed that each user has a different content preference. 
Therefore, a random content preference order and a random skewness are chosen for each user.  
While the content order is chosen using random permutation, the parameter, $\gamma$, is chosen following $\mathrm{Uniform}$ random distribution within a range of maximum $\gamma^{max}$ and minimum $\gamma^{min}$ values. 
Without any loss of generality, the probability that user $u_0$ requests for content $f_k$ is denoted by $\rho_{f_k}^{u_0}$. 
This is modeled based on the $\mathrm{Zipf}$ distribution.

\subsection{Proposed Content Access Protocol}
\label{Content_Access_Protocol}
For accessing the contents, the following practical cases are considered. 

\textbf{Case 1 - \textbf{Local/self cache hit}}: If a tagged user requests the content that is previously cached, the user can directly access the content from its own storage.

\textbf{Case 2 - \textbf{D2D cache hit}}: 
If the required content is not stored in its own storage,  the tagged user sends the content request to the neighboring D2D nodes. 
If any of the neighbors has the content, the user can extract the content from that neighboring user.

\textbf{Case 3 -  \textbf{sBS cache hit}}: 
If the tagged user is under the communication range of any sBS, it maintains its communication via the tagged sBS. 
In this particular case, we have the following sub-cases: 

\textit{Case 3.1}: If the requested content is in the tagged sBS cache, it can access the content directly from there. 
We denote this case as a direct cache hit from the tagged sBS.
	
\textit{Case 3.2}: If the content is not stored in the tagged sBS cache but is available in one of the neighboring sBSs, the tagged sBS extracts the content from the neighboring sBS via its FD capability and delivers it to the tagged user.
We denote this term as soft-sBS (SsBS) cache hit.

\textit{Case 3.3}: If the requested content is not available in any of the sBSs, the tagged sBS forwards the request to the serving MBS. 
If the content is in the serving MBS, it is delivered to the tagged sBS and then to the user. 
This case is denoted as the sBS-MBS cache hit. 
	
\textit{Case 3.4}: If all of the above sub-cases fail, then the MBS extracts the content from the cloud using its FD capability. 
The sBS extracts the content from the MBS using its own FD capability and delivers it to the tagged user. 
This case is denoted as the sBS cache miss.

\textbf{Case 4 - \textbf{MBS cache hit}}: 
If the tagged user is not in the communication range of any of the sBSs, it has to rely on the serving MBS for its communication. 
In this case, we consider the following sub-cases: 

\textit{Case 4.1}: If the requested content is available in the MBS cache, the content is directly delivered to the tagged user. 
This case is denoted as an MBS cache hit.

\textit{Case 4.2}: If the content is not available in the MBS storage and the above case fails, the MBS extracts the content from the cloud using its FD capability.  Then, the content is directly delivered to the user. 
This case is referred as an MBS cache miss.

Without loss of generality, \textit{Case 3} and \textit{(Case 4)} are denoted by the indicator function $\mathbb{I}_s$ and $\mathbb{I}_{m}$, respectively. 
Note that, in \textit{Case 3}, if the tagged user is in the communication ranges of multiple sBSs, it gets connected to the one that provides the best received power.

\section{Edge Caching: Cache Hit Ratio Analysis}	
\label{FD_EDGE_Cost_Analysis}
In this section, we analyze and calculate the local cache hit probabilities.

\subsection{Caching Probabilities}
We now analyze the cache hit probability at different nodes for the cases mentioned in Section \ref{Content_Access_Protocol}. 
Note that a cache hit occurs at a node, if a requested content is available in that node.

\subsubsection{Case 1 - Local/self cache hit} 
The local cache hit probability is denoted as $\mathrm{P_o^u} = \eta_{f_k}^{u_0}$, i.e. the probability of storing the content $f$ at the self cache storage of the tagged user. 

\subsubsection{Case 2 - D2D cache hit} 
The cache hit probability for the D2D nodes can be calculated as follows:
\begin{equation}
\label{Case_2}
	\mathrm{P}_{d}^{u} = \left(1-\eta_{f_k}^{u_0}\right) \left[1 -\prod_{u_i \in \Phi_{u} \backslash u_0 }\left(1 - \eta_{f_k}^{u_i} \right)\right],
\end{equation}
where $\prod_{u_i \in \Phi_{u}}\left(1 - \eta_{f_k}^{u_i} \right)$ means that none of the $\Phi_{u}$ active neighbors (D2D nodes) in its communication range have the content. 
Thus, the complement of that is the probability that at least one of the users stores the content.

\subsubsection{Case 3 - sBS cache hit}
In this case, we calculate the cache hit probabilities achieved via the tagged sBS for the respective sub-cases.

\textit{Case 3.1:} At first, the probability of getting a requested content from the tagged sBS is calculated as follows:
\begin{equation}
\label{Case_3.1}
	\mathrm{P}_{b_o}^{u} = \left(1-\eta_{f_k}^{u_0}\right) \prod_{u_i \in \Phi_{u}\backslash u_0} \left(1 - \eta_{f_k}^{u_i} \right) \eta_{f_k}^{b_0}.
\end{equation}

\textit{Case 3.2:}  The probability of getting a requested content from one of the neighbor sBSs is considered in this sub-case. 
Essentially, this case states that a cache miss has occurred at the tagged sBS.
Mathematically, we express this as follows:
\begin{equation}
\label{Case_3.2}
\begin{aligned}
	\mathrm{P}_{B}^{u} &= \left(1-\eta_{f_k}^{u_0} \right) \prod_{u_i \in \Phi_{u}\backslash u_0 } \left(1 - \eta_{f_k}^{u_i} \right) \left(1 - \eta_{f_k}^{b_0}\right) \\
	&\qquad \qquad \left(1 - \prod_{b_j \in 		\Phi_{b} \backslash b_0}\left( 1 - \eta_{f_k}^{b_j}\right) \right),
\end{aligned}
\end{equation}
where $\Phi_{b}$ is the set of active neighboring sBSs that are in the communication range of the tagged sBS.

\textit{Case 3.3:} 
If sub-case 3.1 and 3.2 fail, the content request is forwarded to the serving MBS via the tagged sBS. 
The cache hit probability, for this case, is calculated as follows:
\begin{equation}
\label{Case_3.3}
\begin{aligned} 
	\mathrm{P}_{M_{\mathbb{I}_s}}^{u} &= \left(1-\eta_{f_k}^{u_0}\right) \prod_{u_i \in \Phi_{u} \backslash u_0} \left(1 - \eta_{f_k}^{u_i} \right) \left(1 - \eta_{f_k}^{b_0}\right) \\
	&\qquad \qquad \prod_{b_j \in \Phi_{b} \backslash b_o} \left( 1 - \eta_{f_k}^{b_j}\right) \eta_{f_k}^{m_0}.
\end{aligned}
\end{equation}

When $\mathbb{I}_s = 1$ - the tagged user is in the communication range of at least one of the sBS, from the above cases and sub-cases, we calculate the total cache hit probability as follows:
\begin{equation}
\label{totalCHR_I_S_1}
\begin{aligned}
	\mathrm{P}_{l}^{\mathbb{I}_s} &
	= 1 - \Bigg[ \left(1 - \eta_{f_k}^{u_0} \right) \prod_{u_i \in \Phi_{u}\backslash u_0} \left(1 - \eta_{f_k}^{u_i}\right)    \left(1 - \eta_{f_k}^{b_0} \right) \\
	&\qquad\qquad \prod_{b_j \in \Phi_{b}\backslash b_0} \left(1 - \eta_{f_k}^{b_j}\right)\Bigg]  \left(1 - \eta_{f_k}^{m_0} \right).   
\end{aligned}
\end{equation}

\textit{Case 3.4:} Now, if the content is not even stored in the MBS cache store, it has to be downloaded from the cloud. 
This case is termed as a cache miss via both sBS and MBS. 
In this case, the MBS initiates its FD mode and download the content from the cloud. 
Therefore, the cache miss probability is calculated from (\ref{totalCHR_I_S_1}) as follows:
\begin{equation}
\label{CacheMiss_IS}
\begin{aligned} 
	\mathrm{P}_{C_{\mathbb{I}_s}}^u \! &
	 =  \Bigg[ \left(1 - \eta_{f_k}^{u_0} \right) \prod_{u_i \in \Phi_{u}\backslash u_0} \left(1 - \eta_{f_k}^{u_i}\right)    \left(1 - \eta_{f_k}^{b_0} \right) \\
	 &\qquad \qquad \prod_{b_j \in \Phi_{b}\backslash b_0} \left(1 - \eta_{f_k}^{b_j}\right)\Bigg]  \left(1 - \eta_{f_k}^{m_0} \right).
\end{aligned} 
\end{equation}

\subsubsection{Case 4 - MBS cache hit} 
Recall that \textit{Case 4} is only considered when the tagged user is not under the coverage region of any of the sBSs.  
First, we consider \textit{Case 4.1} - the requested content is available in the MBS cache (i.e. $\mathbb{I}_m=1$ and $\mathbb{I}_s=0$).
In this sub-case, we calculate the cache hit probability as follows:
\begin{equation}
\label{Case_4.1}
	\mathrm{P}_{M_{\mathbb{I}_M}}^u = \left(1-\eta_{f_k}^{u_0}\right) \prod_{u_i \in \Phi_{u}\backslash u_i } \left(1 - \eta_{f_k}^{u_i} \right) \eta_{f_k}^{m_0}.
\end{equation}
Furthermore, we calculate the total local cache hit probability in this case as follows:
\begin{equation}
\label{Local_CacheHit_Im}
\begin{aligned}
	\mathrm{P}_{l}^{\mathbb{I}_m} 
	&= 1 - \left(1 - \eta_{f_k}^{u_0} \right) \prod_{u_i \in \Phi_{u} \backslash u_0 } \left(1 - \eta_{f_k}^{u_i}\right) \left( 1 - \eta_{f_k}^{m_0}\right). 
\end{aligned}
\end{equation}
Note that we derive the cache miss probability of \textit{Case 4.2} as follows:
\begin{equation}
\label{CacheMiss_C4.2}
	\mathrm{P}_{C_{\mathbb{I}_m}}^u = \left(1 - \eta_{f_k}^{u_0} \right) \prod_{u_i \in \Phi_{u} \backslash u_0 } \left(1 - \eta_{f_k}^{u_i}\right) \left( 1 - \eta_{f_k}^{m_0}\right).
\end{equation}

\section{Edge Caching: Cache Hit Ratio Analysis}
We determine CHR, follwed by successful transmission probabilities in this section.

\subsection{Cache Hit Ratio}
We define CHR as the percentage of the served requests of a requester node from the local nodes. 
In other words, CHR defines the fraction of the requests that are served locally without reaching the cloud.
Let us denote the $\alpha$ portion of the users by the set of $\mathcal{U}_0$. 
Recall that $\rho_{f_k}^{u_0}$ denotes the probability that the tagged user $u_0$ request content $f_k$.
As such, in a heterogeneous caching placement, we determine the fraction of requests of $u_0$ that are served from the local nodes as follows:
\begin{equation}
\label{Cell_CHR}
\begin{aligned}
	\mathrm{CHR} &= \sum_{k=1}^{F} \rho_{f_k}^{u_0} \Bigg[ \eta_{f_k}^{u_0} + \mathrm{P}_d^u \mathrm{P}_{s,f}^{u}  + \underbrace{\Big( \mathrm{P}_{M_{\mathbb{I}_M}}^u \mathrm{P}_{s,f, \mathbb{I}_m =1}^{m_0} \Big)}_{\text{cache hit in case 4}}  \mathbb{I}_m + \\
	&\qquad \qquad \quad \underbrace{\Big(  \mathrm{P}_{b_0}^{u} \mathrm{P}_{s,f}^{b_0} + \mathrm{P}_B^u \mathrm{P}_{s,f}^{b} + \mathrm{P}_{M_{\mathbb{I}_s}}^u \mathrm{P}_{s,f}^{m_0} \Big)}_{\text{cache hit in case 3}} \mathbb{I}_s \Bigg],
\end{aligned}
\end{equation}
where the first term represents the self cache hit, while the second term represents the successfully achieved cache hit from D2D neighbors. 
The contents inside $\left(\cdot\right)$ in the third term and in the fourth term are the successfully achieved cache hit from \textit{Case 3} and \textit{Case 4}, respectively.
Moreover, $\mathrm{P}_{s,f}^{*}$ represents the successful transmission probability for the respective * cases.
Note that the transmission success probability between two nodes does not depend on the content index. 
Therefore, we mention the success probability as $\mathrm{P_{s,f}^{*}}$ instead of $\mathrm{P_{s,f_k}^{*}}$.

\subsection{Probability of Successful Transmission}

Now, we calculate the transmission success probabilities among different nodes. 
When a tagged user request a content, interference comes from - other active D2D users, active sBSs and MBS.
The wireless channel between two nodes follows a Rayleigh fading distribution with $\mathcal{CN}(0,1)$. 
Let us denote the channel between node $i$ and node $j$ by $h_{ij}$.  
Let us also denote the threshold SINR for successful communication by $\phi$ dB. 
The transmission power of the user, sBS and MBS are denoted by $p_u$, $p_b$ and $p_m$, respectively. 
Moreover, the path loss exponent is denoted by $\beta$.

Now, let $\gamma_i^j$, $d_i^j$ and $I_{ij}$ denote the SINR at node $i$ served from node $j$, distance between the nodes and total interference at node $i$, respectively. 
We then derive the SINR values for different cases and sub-cases in equation (\ref{SINR_Cal}).
Owing to the space constraint, the detail derivations of these probabilities are omitted.
However, the final tight closed form approximations are provided in equation (\ref{P_sf}-\ref{A_B}).
Also, note that we do not consider the case of obtaining the content from the cloud, when we calculate $\mathrm{CHR}$. 
This is due to the fact that we are interested in calculating the percentage of served request from the local nodes only.

\section{Cache Hit Ratio Maximization using Particle Swarm Optimization}
\label{Opt_Prop_Algo}
We present our objective function, followed by the proposed M-PSO algorithm in this section.

\subsection{CHR Maximization Objective Function}
To this end, we calculate the average cache hit ratio for the requesting nodes, which is denoted by $\Sigma$.
The detail derivation of the $\Sigma$ is shown in (\ref{TCHR}).
Our objective is to maximize the $\Sigma$ given that the storage constraints are not violated. 
Thus, we express the objective function in heterogeneous caching model case as follows: 
\begin{subequations}
	\begin{align}
	\mathbf{P_1:}\quad	&  \underset{\eta_{f_k}^{u_i}, \eta_{f_k}^{b_j}, \eta_{f_k}^{m_l}}  {\text{maximize}} \quad \Sigma \\
	& \quad \text{s. t.} ~\label{CHR_c1}  \sum_{k=1}^{F} \eta_{f_k}^{u_i} \leq \mathcal{C}_u, \quad \forall~ u_i \in \{\mathcal{U}\}, f_k \in \{\mathcal{F}\} \\ 
	& \label{CHR_c2} \qquad 	\sum_{k=1}^{F} \eta_{f_k}^{b_j} \leq \mathcal{C}_b, \quad \forall~ b_j \in \{\mathcal{B}\}, f_k \in \{\mathcal{F}\}\\
	& \label{CHR_c3} \qquad 	\sum_{k=1}^{F} \eta_{f_k}^{m_l} \leq \mathcal{C}_m, \quad \forall ~ m_l \in \{\mathcal{M}\}, f_k \in \{\mathcal{F}\} \\
	& \label{CHR_c4} \qquad  	0 \leq \eta_{f_k}^{u_i} \leq 1,~ 0 \leq \eta_{f_k}^{b_j} \leq 1,~ 0 \leq \eta_{f_k}^{m_l} \leq 1,	
	\end{align}
	\label{opt_TCHR}
\end{subequations}
where the constraints in (\ref{CHR_c1}-\ref{CHR_c3}) ensure the physical storage size limitations of the user, the sBS and the MBS, respectively, while the constraints in (\ref{CHR_c4}) are due to the probability range in $[0,1]$.

We intend to find optimal caching placements variables that deliver us the optimal solutions.
The motivation of $\mathbf{P}_1$ is to ensure that the requested contents are delivered locally instead of overwhelming the core network during busy traffic hours.
However, in general, problem $\mathbf{P_1}$ is non-convex \cite{8667875} by nature and may not be solved efficiently in a polynomial-time due to the nonlinear and combinatorial content placement variables. 
Had we have binary decision parameters, it is not hard to see that the $\mathbf{P}_1$ would have been reduced to a Knapsack problem, which is widely recognized as an NP-complete problem.
Nevertheless, each of our decision variables is a probability that is in $[0,1]$.
There is an infinite number of possible values, to determine the optimal solution from, in this range.
Therefore, the use of typical metaheuristic solutions such as genetic algorithms may not be a suitable choice.
Thanks to particle swarm optimization (PSO) technique, we can leverage its fundamental concept to get to a modified version of it that is suitable for a complex combinatorial problem such as $\mathbf{P}_1$.
We discuss our proposed modified PSO (M-PSO) framework in what follows.

\begin{figure*}[!t] 
\begin{subequations}
\label{SINR_Cal}
\begin{align}
    \gamma_{u_0}^u &= \frac{p_u h_{u_0 u_t} d_{u_0 u_t}^{-\beta}}{\sigma^2 + I_{u_0 u}}, \quad 
    \gamma_{u_0}^{b_0} = \frac{p_b h_{u_0 b_0} d_{u_0 b_0}^{-\beta}}{\sigma^2 + I_{u_0 b_0}}, \quad
    \gamma_{b_0}^{b} = \frac{p_b h_{b_0 b_t} d_{b_0 b_t}^{-\beta}}{\sigma^2 + I_{b_0 b}}, \quad
    \gamma_{b_0}^{m_0} = \frac{p_m h_{b_0 m_0} d_{b_0 m_0}^{-\beta}}{\sigma^2 + I_{b_0 m_0}}, \quad
    \gamma_{u_0}^{m_0} = \frac{p_m h_{u_0, m_0} d_{u_0,m_0}^{-\beta}}{\sigma^2 + I_{u_0, m_0}},
\end{align}
\end{subequations}
where the interference, $I_{ij}$s, are calculated as follows: 
\begin{subequations}
\begin{align}
    I_{u_0 u}& = \sum_{u \in \Phi_{u} \backslash \{u_0,u_t\} } p_u h_{u_0 u} d_{u_0 u}^{-\beta} + \sum_{b_l \in \{\mathcal{B}\}_{b_0}^{B}} p_{b_l} h_{u_0 b_l} d_{u_0 b_l}^{-\beta} \mathbb{I}_{b_l} + \sum_{b_l \in \{\mathcal{B}\}_{b_0}^{B}, ~u_l \in \Phi_{u} \backslash{\{u_0, u_t\}}} p_m h_{u_0 m} d_{u_0 m}^{-\beta} \mathbb{I}_{m_0} \\
    I_{u_0 b_0} &=  \sum_{u \in \Phi_{u} \backslash u_0} p_u h_{u_0 u} d_{u_0 u}^{-\beta} +  \sum_{b_l \in \{\mathcal{B}\}_{b_l}^{B} \backslash b_0}  p_{b_l} h_{u_0 b_l} d_{u_0 b_l}^{-\beta} \mathbb{I}_{b_l}  +  \sum_{b_l \in \{\mathcal{B}\}_{b_l}^{B}\backslash b_0,~ u_l \in \Phi_{u} \backslash u_0 } p_m h_{u_0 m} d_{u_0 m}^{-\beta} \mathbb{I}_{m_0}\\
    I_{b_0 b} &= \sum_{u \in \Phi_{u}\backslash u_0} p_u h_{b_0 u} d_{b_0 u}^{-\beta} + \sum_{b \in \{\mathcal{B}\}\backslash \{b_0, b_t\}} p_b h_{b_0 b} d_{b_0 b}^{-\beta} +  h_{b_0 b_0} \zeta p_b + \sum_{b_l \in \{\mathcal{B}\}_{b_l}^{B}\backslash\{b_0, b_t\},~ u \in \Phi_{u} \backslash u_0 } p_m h_{u_0 m} d_{u_0 m}^{-\beta} \mathbb{I}_{m_0}\\
    I_{b_0 m_0} &= \sum_{u \in \Phi_{u} \backslash u_0} p_u h_{b_0 u} d_{b_0 u}^{-\beta} + \sum_{b \in \{\mathcal{B}\}\backslash b_0} p_b h_{b_0 b} d_{b_0 b}^{-\beta} + h_{b_0 b_0} \zeta p_b + \sum_{b_l \in \{\mathcal{B}\}_{b_l}^{B}\backslash b_0, u \in \Phi_{u}\backslash u_0 } p_m h_{u_0 m} d_{u_0 m}^{-\beta} \mathbb{I}_{m_0}\\
    I_{u_0,m_0} & = \sum_{u \in \Phi_{u} \backslash u_0 } p_u h_{u_0 u} d_{u_0 u}^{-\beta}  + \sum_{b_l \in \{\mathcal{B}\},~ u_l \in \Phi_{u} \backslash u_0}  p_m h_{u_0 m} d_{u_0 m}^{-\beta} \mathbb{I}_{m_0}.
\end{align}
\end{subequations}
\begin{subequations} 
\label{P_sf}
\begin{align}
    \mathrm{P}_{s,f}^{u}& = \frac{A}{B} \left[1 - \exp\left(-\pi R_u^2 B\right) \right],\qquad
    \mathrm{P}_{s,f}^{b_0} = \frac{A_1}{B_1} \left[ 1 - \exp(-\pi R_b^2 B_1)\right], \qquad
    \mathrm{P}_{s,f, \mathbb{I}_m =1}^{m_0} = \frac{A_2}{B_2} \left[ 1 - \exp(-\pi R_m^2 B_2)\right],\\
    \mathrm{P}_{s,f}^b  &= \Bigg[ \int_{r>0} \Bigg\{ f_{d_1}(r) \exp \Bigg(\frac{- \pi \alpha \lambda_u  \left( \phi \frac{p_u}{p_b} \right)^{\frac{2}{\beta}} r^2 } {\sinc(\frac{2}{\beta})} \Bigg)  \exp \Bigg(\frac{- \pi \lambda_b  \phi^{\frac{2}{\beta}} r^2 } {\sinc(\frac{2}{\beta})} \Bigg) \exp  \left(-\phi r^{\beta} \bar{\zeta} \right) \exp \Bigg(\frac{- \pi \lambda_m  \left( \phi \frac{p_m}{p_b}\right)^{\frac{2}{\beta}} r^2 } {\sinc(\frac{2}{\beta})} \Bigg) \Bigg\} dr \Bigg] \nonumber \\ &\qquad \Bigg\{\frac{A_1}{B_1} \left[ 1 - \exp(-\pi R_b^2 B_1)\right] \Bigg\},\\
    \mathrm{P}_{s,f}^{m_0} & = \Bigg[ \int_{r>0} \Bigg\{ f_{d_2}(r)  \exp \Bigg(\frac{- \pi \alpha \lambda_u  \left( \phi \frac{p_u}{p_m} \right)^{\frac{2}{\beta}} r^2 } {\sinc(\frac{2}{\beta})} \Bigg) \exp \Bigg(\frac{- \pi \lambda_b  \left( \phi \frac{p_b}{p_m} \right)^{\frac{2}{\beta}} r^2 } {\sinc(\frac{2}{\beta})} \Bigg) \exp  \left(-\phi r^{\beta} \bar{\zeta} \right) \exp \Bigg(\frac{- \pi \lambda_m r^2 \phi^{\frac{2}{\beta}} } {\sinc(\frac{2}{\beta})} \Bigg) \Bigg\} dr \Bigg] \nonumber \\ 
    & \qquad \Bigg\{ \frac{A_1}{B_1} \left[ 1 - \exp(-\pi R_b^2 B_1)\right] \Bigg\},
\end{align}
\end{subequations}
where $\bar{\zeta}$ is the self-interference \cite{tan2018d2d} due to FD communication. Moreover, $A$, $B$, $A_1$, $B_1$, $A_2$ and $B_2$ are calculated as follows:
\begin{subequations}
\label{A_B}
\begin{align}
    A &= \frac{(1-\alpha)\lambda_u}{1- \exp \left[-\pi (1-\alpha) \lambda_u R_u^2 \right]}, \qquad
    B = \lambda_u \left( (1-\alpha) + \frac{\alpha  \phi^{\frac{2}{\beta}}}{\sinc(2/\beta)} \right)  + \frac{ \lambda_b  \left( \phi \frac{p_b}{p_u} \right)^{\frac{2}{\beta}}} {\sinc(\frac{2}{\beta})} + \frac{ \lambda_m  \left( \phi \frac{p_m}{p_u} \right)^{\frac{2}{\beta}}} {\sinc(\frac{2}{\beta})}, \\
    A_1 &= \frac{\lambda_b}{1 - \exp (\pi \lambda_b R_b^2)}, \qquad
    B_1 = \lambda_b \left[ 1 + 2 \phi^{\frac{2}{\beta}} \int_{\phi^{\frac{-2}{\beta}} }^\infty \left( \frac{1}{1 + u^{\frac{\beta}{2}}}\right) du \right] + \frac{\alpha \lambda_u  \left( \phi \frac{p_u}{p_b}\right)^{\frac{2}{\beta}}}{\sinc \left(\frac{2}{\beta}\right)} + \frac{ \lambda_m  \left(\phi \frac{p_m}{p_b} \right)^{\frac{2}{\beta}}}{\sinc \left( \frac{2}{\beta}\right)}, \\
    A_2 &= \frac{\lambda_m}{1 - \exp (\pi \lambda_m R_m^2)}, \qquad 
    B_2 = \lambda_m \left[  1 + \frac{\phi ^{\frac{2}{\beta}}}{\sinc \left( \frac{2}{\beta}\right)} \right] + \frac{\alpha \lambda_u  \left( \phi \frac{p_u}{p_m}\right)^{\frac{2}{\beta}}}{\sinc \left(\frac{2}{\beta}\right)}.
\end{align}
\end{subequations}
\begin{equation}
\label{TCHR}
\begin{aligned}
    \Sigma &= \frac{1}{|\mathcal{U}_0|} \sum_{ u_0 \in \mathcal{U}_0} \sum_{k=1}^{F} \rho_{f_k}^{u_0} \Bigg\{ \eta_{f_k}^{u_0} +  \left(1-\eta_{f_k}^{u_0}\right) \left[1 -\prod_{u_i \in \Phi_{u} \backslash u_0 } \left(1 - \eta_{f_k}^{u_i} \right)\right] \mathrm{P}_{s,f}^{u} +  \Bigg( \left(1-\eta_{f_k}^{u_0}\right) \prod_{u_i \in \Phi_{u}\backslash u_0} \!\!\!\! \!\! \left(1 - \eta_{f_k}^{u_i} \right) \eta_{f_k}^{b_0} \mathrm{P}_{s,f}^{b_0} + \\
    &\qquad \left(1-\eta_{f_k}^{u_0} \right) \! \! \!\! \prod_{u_i \in \Phi_{u}\backslash u_0 } \! \! \!\! \left(1 - \eta_{f_k}^{u_i} \right) \!\! \left(1 - \eta_{f_k}^{b_0}\right) \!\! \left(1 - \!\!\!\prod_{b_j \in \Phi_{b} \backslash b_0} \!\!\!\! \left( 1 - \eta_{f_k}^{b_j}\right) \!\!\! \right) \mathrm{P}_{s,f}^{b} + \left(1-\eta_{f_k}^{u_0}\right) \prod_{u_i \in \Phi_{u} \backslash u_0} \left(1 - \eta_{f_k}^{u_i} \right) \left(1 - \eta_{f_k}^{b_0}\right) \\
    &\qquad \prod_{b_j \in \Phi_{b} \backslash b_o} \left( 1 - \eta_{f_k}^{b_j}\right) \eta_{f_k}^{m_0}  \mathrm{P}_{s,f}^{m_0} \Bigg) \mathbb{I}_s +  \Bigg( \left(1-\eta_{f_k}^{u_0}\right) \prod_{u_i \in \Phi_{u}\backslash u_i } \left(1 - \eta_{f_k}^{u_i} \right) \eta_{f_k}^{m_0} \mathrm{P}_{s,f, \mathbb{I}_m =1}^{m_0} \Bigg)  \mathbb{I}_m  \Bigg\} .
\end{aligned}
\end{equation}
\end{figure*}

\subsection{Modified-Particle Swarm Optimization Algorithm}
PSO is a swarm intelligence approach that guarantees to converge \cite{kennedy2010particle}. 
In this meta-heuristic algorithm, all possible sets of candidate solutions are named as the particles - denoted by $i$. 
Each particle has a position - denoted by $x_i$.
Furthermore, it maintains a personal best position of each particle and the global best positions of the entire swarm. 
These two terms are denoted by $p_i^{best}$ and $g^{best}$, respectively. 
The algorithm evolves, with an exploration and exploitation manner, by adding a velocity term - $v_i^t$ at each particle's previous position aiming to converge at the global optima. 
The following two simple equations, thus, govern the PSO algorithm.
\begin{align}
    v_i^{t+1} &= av_i^t + \psi_1 \epsilon_1 \left(p_i^{best} - x_i\right) + \psi_2 \epsilon_2 \left(g^{best} - x_i\right), \label{velocity_gen}\\
    x_i^{t+1} &= x_i^t + v_i^{t}, \label{position_gen}
\end{align}
where $a$, $\psi_1$ and $\psi_2$ are the parameters that needs to be selected properly. 
Moreover, $\epsilon_1$ and $\epsilon_2$ are two $\mathrm{Unifrom}$ random variables. 
Note that $\psi_1$ and $\psi_2$ are positive acceleration coefficients, which are also known as the cognitive and social learning factors \cite{8667875}, respectively.
While this is a general framework for the PSO algorithm, it may not be used directly in constraint optimization \cite{hu2003engineering}. 
In our objective function, each particle must have a position matrix - each dimension of which must not violate the restrictions. 
Therefore, in the following, we modify the PSO algorithm to solve our optimization problem efficiently.

Let $P$ be numbers of particles. 
Let $\pmb{\eta}_{f}^{u_i}$ denote the caching probabilities of user $u_i$ for all contents $f_k \in \{\mathcal{F} \}$. 
Then, this parameter has a size of $F \times 1$. 
Similarly, for all sBS and MBS, let $\pmb{\eta}_{f}^{b_j}$ and $\pmb{\eta}_{f}^{m_l}$ denote their caching placement probabilities for all contents. 
Then, all of these parameters can be stacked into a matrix with dimension of $(|\mathcal{U}| + |\mathcal{B}| + |\mathcal{M}|) \times |\mathcal{F}|$, which is the exact shape of each particle. 
Let the current position of each of these particles be denoted by $\mathbf{X}_i^{t}$. 
Note that in this case, each particle's position $\mathbf{X}_i^{t}$ has a shape of $(|\mathcal{U}| + |\mathcal{B}| + |\mathcal{M}|) \times |\mathcal{F}|$. 
Let $\mathbf{V}_i^{t} \in \mathbb{R}^{(|\mathcal{U}| + |\mathcal{B}| + |\mathcal{M}|) \times |\mathcal{F}|}$ denote the velocity. Furthermore, the personal best position of particle $i$ is denoted by $\mathbf{P}_i^{\mathrm{best}}$, while the global best for the entire swarm is denoted by $\mathbf{G}^{\mathrm{best}}$. 
Therefore, each particle updates its velocity with social and individual cognition. 
We use the following equation to govern these updates. 
\begin{equation}
\small
\begin{aligned}
\label{Velocity}
    \!\!\mathbf{V}_{i}^{t+1} \!\!&=\! a\mathbf{V}_{i}^t + \psi_1 \left[\pmb{\mathcal{E}}_1 \! \odot\!\! \left(\mathbf{P}_{i}^{\mathrm{best}}\!\! - \! \mathbf{X}_{i}^{t}\right)\right] \! + \! \psi_2 \left[\pmb{\mathcal{E}}_2 \! \odot \! \! \left(\mathbf{G}^{\mathrm{best}}\!\!\! - \mathbf{X}_{i}^t\! \right)\!\right]\!\!,
\end{aligned} 
\end{equation}
where $a$, $\psi_1$ and $\psi_2$ are the parameters as described in (\ref{velocity_gen}). 
Moreover, $\pmb{\mathcal{E}}_1$ and $\pmb{\mathcal{E}}_2$ are two matrices with sizes of $\mathbb{R}^{(|\mathcal{U}| + |\mathcal{B}| + |\mathcal{M}|) \times |\mathcal{F}|}$, where their element is drawn from $\mathrm{Unifrom}$ random distribution. 
Finally, $\odot$ represents Hadamard product.

The position of each particle is then updated by the velocity similar to (\ref{velocity_gen}). 
However, as we have the constraints as in (\ref{CHR_c1})-(\ref{CHR_c4}), we need to modify this equation accordingly. 
Let $\mathbf{X}_{i_{int}}^{t+1}$ denote an intermediate updated position of particle $i$ as shown in the following expression.
\begin{equation}
\small
\label{Particle_new_position}
    \mathbf{X}_{i_{int}}^{t+1} = \mathbf{X}_{i}^t + \mathbf{V}_{i}^t .
\end{equation}
We consider this intermediate position to keep each particle's position in the feasible search space.  
Besides, we also perform the necessary normalization and scaling.
Note that from this intermediate particle position leads to a normalized particle position.
This parameter is then used as the current particle position $\mathbf{X}_i^{t}$. 
Moreover, the ultimate goal for each particle is to converge at an optimal position $\mathbf{X}_{i}^{*}$ (i.e. the global best $\mathbf{G}^{\mathrm{best}}$).

Algorithm \ref{Collaborative_Edge_Caching_Algorithm_CHR} summarizes the steps of the proposed algorithm. Note that our proposed algorithm can be implemented to solve any similar hard combinatorial problems.

\begin{algorithm} [t!]
\caption{CHR Maximization using M-PSO}
	\begin{algorithmic} [1]
		\For {each particle, $i = 1,2,\dots,P$} 
		\State {$\mathbf{X}_i = [~]$, $\mathbf{V}_i = [~]$}
		\For {each dimension $j = 1,2, \dots, D$}  \Comment{$D=|\mathcal{U}| + |\mathcal{B}| + |\mathcal{M}|$}
		\State initialize the particles positions, $\mathbf{x}_{ji} $ with uniform random vector of size $\mathbb{R}^{|\mathcal{F}|}$ by making sure $\sum_{k = 1}^{F} {x}_{ji}[k] = 1$ and $0 \leq {x}_{ji}[k] \leq \frac{1}{\mathcal{C}^{j}}$, $\forall ~k \in \mathcal{F}$; then set $\mathbf{X}_i [j,:] \leftarrow  \mathbf{x}_{ji}$ \Comment{$\mathcal{C}^j$ is the cache storage of the node in $j^{th}$ dimension} \label{Particle_pos_initialization}
		\State initialize particles velocity, $\mathbf{v}_{ji}$ with uniform random vector of size $\mathbb{R}^{|\mathcal{F}|}$ by making sure $\sum_{k = 1}^{F} {v}_{ji}[k] = 1$ and $0 \leq {v}_{ji}[k] \leq \frac{1}{\mathcal{C}^{j}}$, $\forall ~k \in \mathcal{F}$; then set $\mathbf{V}_i [j,:] \leftarrow  \mathbf{v}_{ji}$ \label{Particle_vel_initialization}
		\EndFor 
		\State set particle best position, $\mathbf{P}_i^{best}$ as the initial position
		\If {$\Sigma \left(\mathbf{P}_{i}^{\mathrm{best}}\right) > \mathrm{\Sigma} \left(\mathbf{G}^{\mathrm{best}}\right)$} \label{cond1}
		\State $\mathbf{G}^{\mathrm{best}} \leftarrow \mathbf{P}_i^{\mathrm{best}}$
		\EndIf
		\EndFor
		\While {termination criteria has not met}
		\For {each particle, $i$}
		\For {each dimension, $j = 1,2, \dots, D$}
		\State draw uniform random vectors, $\pmb{\epsilon}_1$ and $\pmb{\epsilon}_1$ of size $\mathbb{R}^{|\mathcal{F}|}$
		\State set $\mathbf{v}_{ji} \leftarrow a \mathbf{v}_{ji} + \psi_1 \left[\pmb{\epsilon}_1 \odot \left(\mathbf{p}_{ji}^{\mathrm{best}} - \mathbf{x}_{ji}\right) \right] + \psi_2 \left[ \pmb{\epsilon}_2 \odot \left(\mathbf{g}_j^{\mathrm{best}} - \mathbf{x}_{ji}  \right)\right] $ 
		\State set $\mathbf{V}_i [j,:] \leftarrow  \mathbf{v}_{ji} $
		\EndFor
		\State update particles intermediate position, $\mathbf{X}_{i_{int}}$ 
		\State $\mathbf{X}_{i}^{\mathrm{scl}} = [~]$, $\mathbf{P}_{i}^{\mathrm{scl\_best}} = [~]$,  $\mathbf{G}^{\mathrm{scl\_best}} = [~]$
		\For{each dimension  $j = 1,2, \dots, D$ }
		\State $\mathrm{random\_hike} \leftarrow \mathrm{randint}(\mathcal{C}^j)$ \label{random_hike}
		\For {i in $len(\mathrm{random\_hike})$}
		\State $\mathbf{X}_{i_{int}}[j, \mathrm{randint}(F)] \leftarrow \frac{\sum_{k=1}^{F} \mathbf{X}_{i_{int}}[j,:]}{\mathcal{C}^j} $  \label{random_hike_nodes}
		\EndFor
		\State $\mathbf{X}_{i}[j,:] \leftarrow \frac{\mathbf{X}_{i_{int}}[j,:]}{\sum_{k=1}^{F} \mathbf{X}_{i_{int}}[j,:]} $; $\mathbf{X}_{i}^{\mathrm{scl}}[j,:] \leftarrow \mathcal{C}^{j} \mathbf{X}_{i}[j,:] $ \Comment{Normalized particle position}
		\State $\mathbf{P}_{i}^{\mathrm{scl\_best}} [j,:] \leftarrow \mathcal{C}^{j} \mathbf{P}_{i}^{\mathrm{best}}[j,:] $ \label{P_best_Scalling}
		\State $\mathbf{G}_{i}^{\mathrm{scl\_best}} [j,:] \leftarrow \mathcal{C}^{j} \mathbf{G}_{i}^{\mathrm{best}}[j,:] $ \label{G_best_Scalling}
		\EndFor 
		\If{$\Sigma \left(\mathbf{X}_{i}^{\mathrm{scl}}\right) > \mathrm{\Sigma} \left(\mathbf{P}_{i}^{\mathrm{scl\_best}}\right)$} \label{cond2}
		\State $\mathbf{P}_i^{\mathrm{best}} \leftarrow \mathbf{X}_i$
		\State do necessary scaling following step \ref{P_best_Scalling}
		\If{$\Sigma\left(\mathbf{P}_{i}^{\mathrm{scl\_best}} \right) > \mathrm{\Sigma} \left(\mathbf{G}^{\mathrm{scl\_best}}\right)$} \label{cond3}
		\State $ \mathbf{G}^{\mathrm{best}} \leftarrow \mathbf{P}_i^{\mathrm{best}}$
		\EndIf
		\EndIf 
		\EndFor
		\EndWhile
		\State \textbf{return} $ \mathbf{G}^{\mathrm{best}}$ and do necessary scaling following step \ref{G_best_Scalling} and \textbf{return} $\mathbf{G}^{\mathrm{scl\_best}}$
	\end{algorithmic} \label{Collaborative_Edge_Caching_Algorithm_CHR}
\end{algorithm}

We model the algorithm such a way that we deal with the normalized particle position and velocity. 
The constraints guide us to restrict the particle position in a probability range, while the summation cannot exceed the cache storage capacity of the respective node. 
Therefore, we consider to limit the initial values in the range of $[0,~1/\mathcal{C}^j]$.
By doing so, when we perform the necessary scaling, the obtained number does not violate the probability range. 
Then, we correspondingly initialize the particle position and the velocity in steps \ref{Particle_pos_initialization} and \ref{Particle_vel_initialization} following this notion. 
Furthermore, the caching probabilities of the nodes in dimension $j$ are limited to $\mathcal{C}^j$, in steps \ref{random_hike} and \ref{random_hike_nodes}, hence, we choose the random number of contents, $\mathrm{randint(\mathcal{C}^j)} $, to be stored with higher probability values.
We stress the fact that, although our proposed M-PSO is a modified version of PSO, it inherits all properties of the original PSO algorithm. 
As such, it is not hard to analyze the convergence and complexities of our proposed algorithm following the analysis of the original PSO algorithm \cite{985692}.

\section{Results and Discussion}
\label{Result}
For the simulation, user are considered to be distributed in a $2$D plane following a HPPP of intensity, $\lambda_u \in [10^{-4},10^{-3}]$ (per $m^2$). 
The low powered sBS are drawn following another HPPP of intensity, $\lambda_b \in [10^{-6},10^{-5}]$,  (per $m^2$). 
For the MBS, $\lambda_m  = 1.5^{-7}$ (per $m^2$) is considered. 
The coverage radii of the user, sBS and MBS are taken as $R_u = 15 ~m$ , $R_b = 150 ~m$, $R_m = 500 ~m$, respectively. 
Total contents in the catalog $|\mathcal{F}| = [10, 50]$, $\alpha  \in [0.2, 0.5]$ and the skewness, $\gamma$ of the $\mathrm{Zipf}$ distribution is considered to be selected uniformly in between \{0.1, 2.5\}. 
For the M-PSO algorithm, we set $a =0.9$ and $\psi_1 = \psi_2=0.4$. 
Moreover, $p_u = 23$ dBm, $p_b=26$ dBm, $p_m = 43$ dBm, $\phi = 10^{-8}$ dB, $\beta = 4$, $\zeta=0.01$ and $\sigma^2 = -174$ dBm/Hz are considered. 
We apply Monte Carlo simulation methods while performing our evaluation. 
In the following, we use the proposed M-PSO algorithm to attain the optimal caching placement solution. 
After that, we study its performances for our hard-combinatorial maximization problem. 


\subsection{Cache Placement}
\begin{figure*} 
	\centering
    \includegraphics[width=\textwidth] {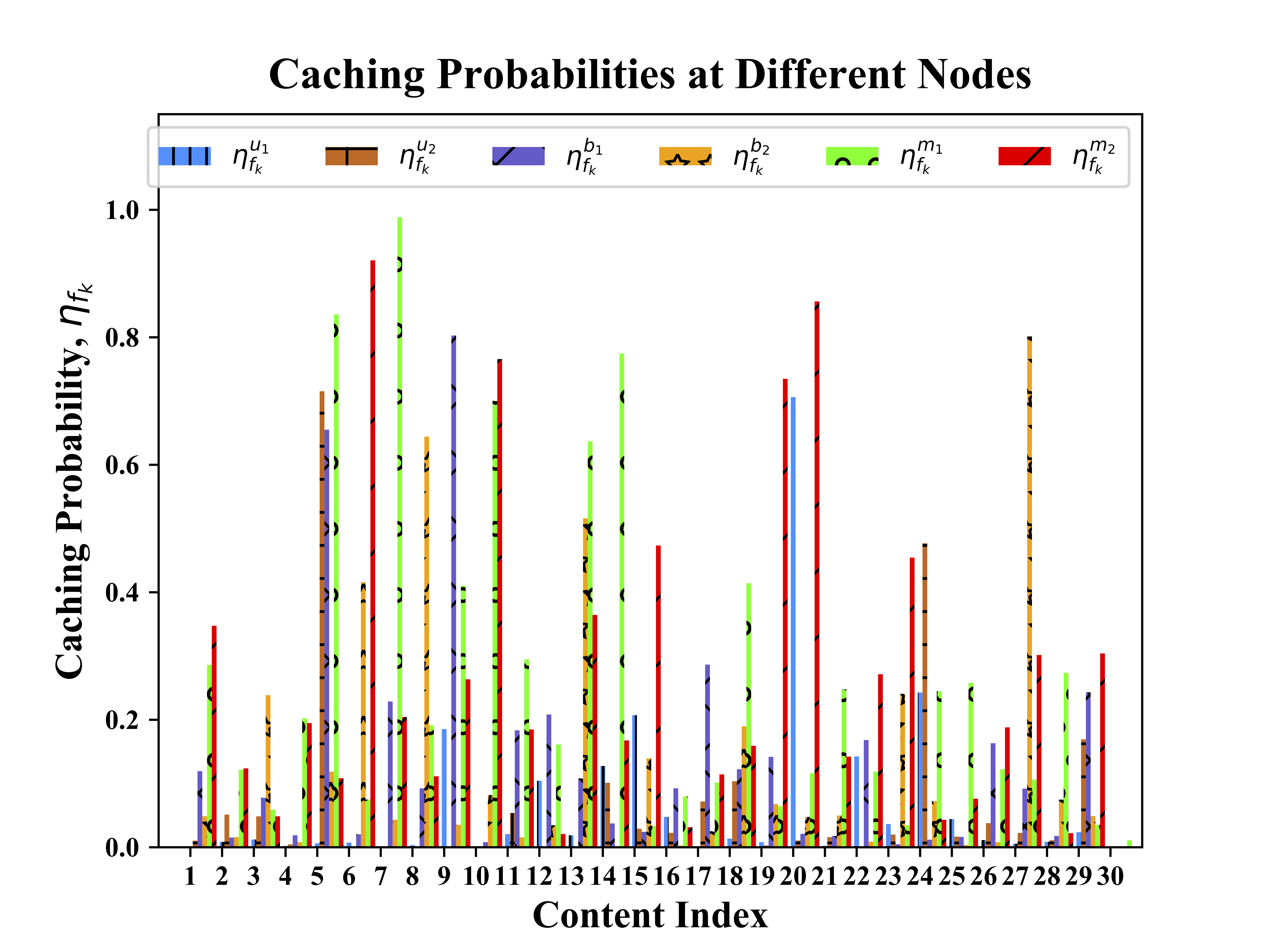}
	\caption{Obtained caching probabilities at the local nodes when $\mathcal{C}_d = 2$, $\mathcal{C}_b = 4$ and $\mathcal{C}_m = 8$}
	\label{Caching_Probabilities}
\end{figure*}

To show the effectiveness of the proposed algorithm, we first validate that the obtained results do not violate any of the constraints. 
The obtained global best $\mathbf{G}^{\mathrm{scl\_best}}$, using Algorithm \ref{Collaborative_Edge_Caching_Algorithm_CHR}, is therefore scrutinized as follows. 
Note that it must not violate any of the caching storage constraints of the edge nodes.
Besides, each of the caching probabilities must be in the range of $[0,1]$.
Furthermore, each node must store different copies of the content. 
Notice that we have applied all of these constraints in our proposed algorithm.
Therefore, it is expected that the obtained results will satisfy these constraints. 
The caching probabilities of the $1^{st}$ and $2^{nd}$ respective D2D users, sBSs and MBSs are illustrated in Fig. \ref{Caching_Probabilities}.
Notice that each node stores different copies. 
Moreover, caching probabilities and storage constraints are also satisfied. 

\begin{figure} 
	\centering
    \includegraphics[width= 0.5 \textwidth] {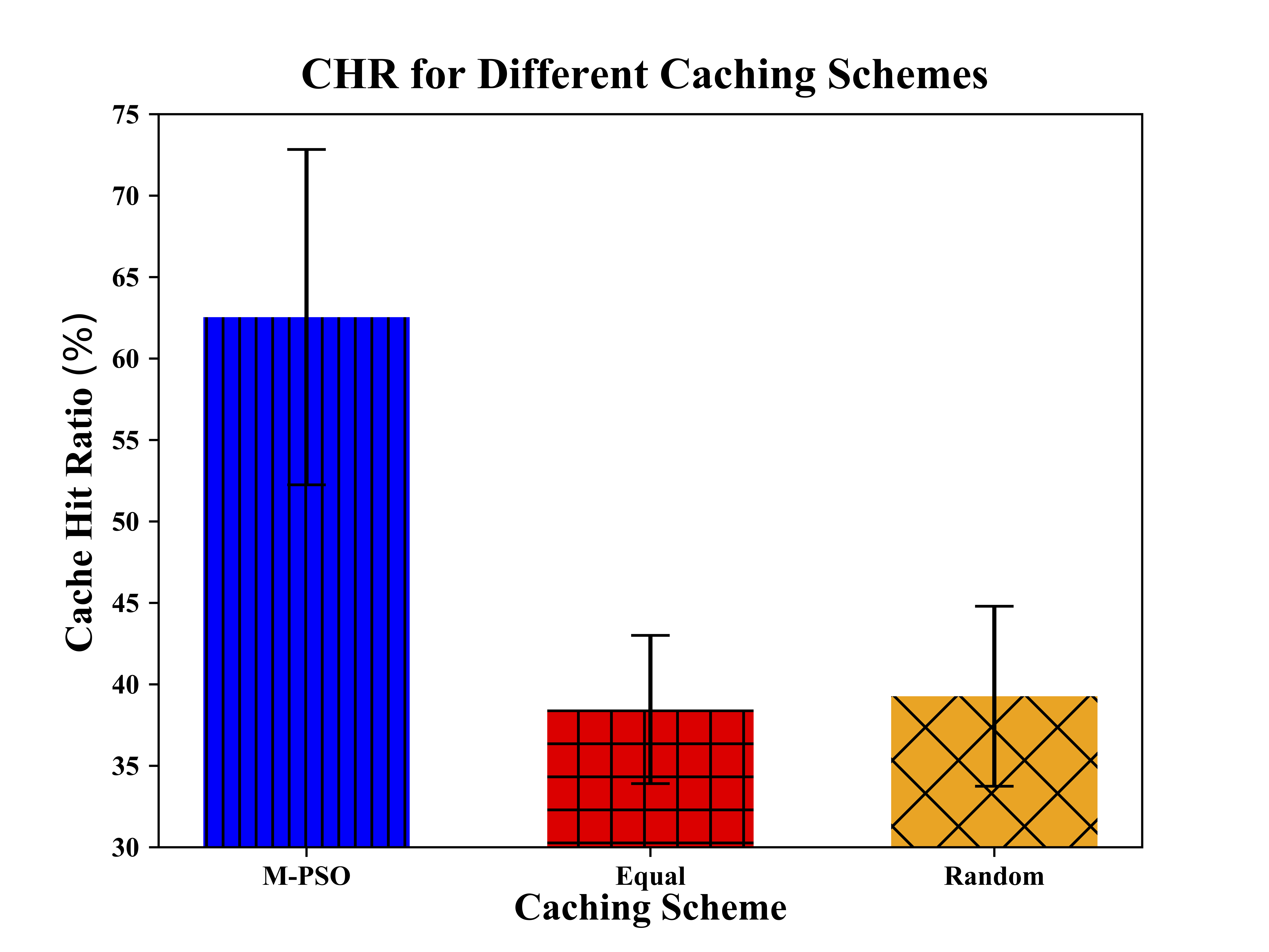}
	\caption{CHR using the proposed M-PSO algorithms for $100$ iteration, $|\mathcal{F}| = 30$, $\mathcal{C}_d = 2$, $\mathcal{C}_b = 4$ and $\mathcal{C}_m = 8$}
	\label{Comparision}
\end{figure}

\begin{figure}
\centering
    \includegraphics[width= 0.5 \textwidth] {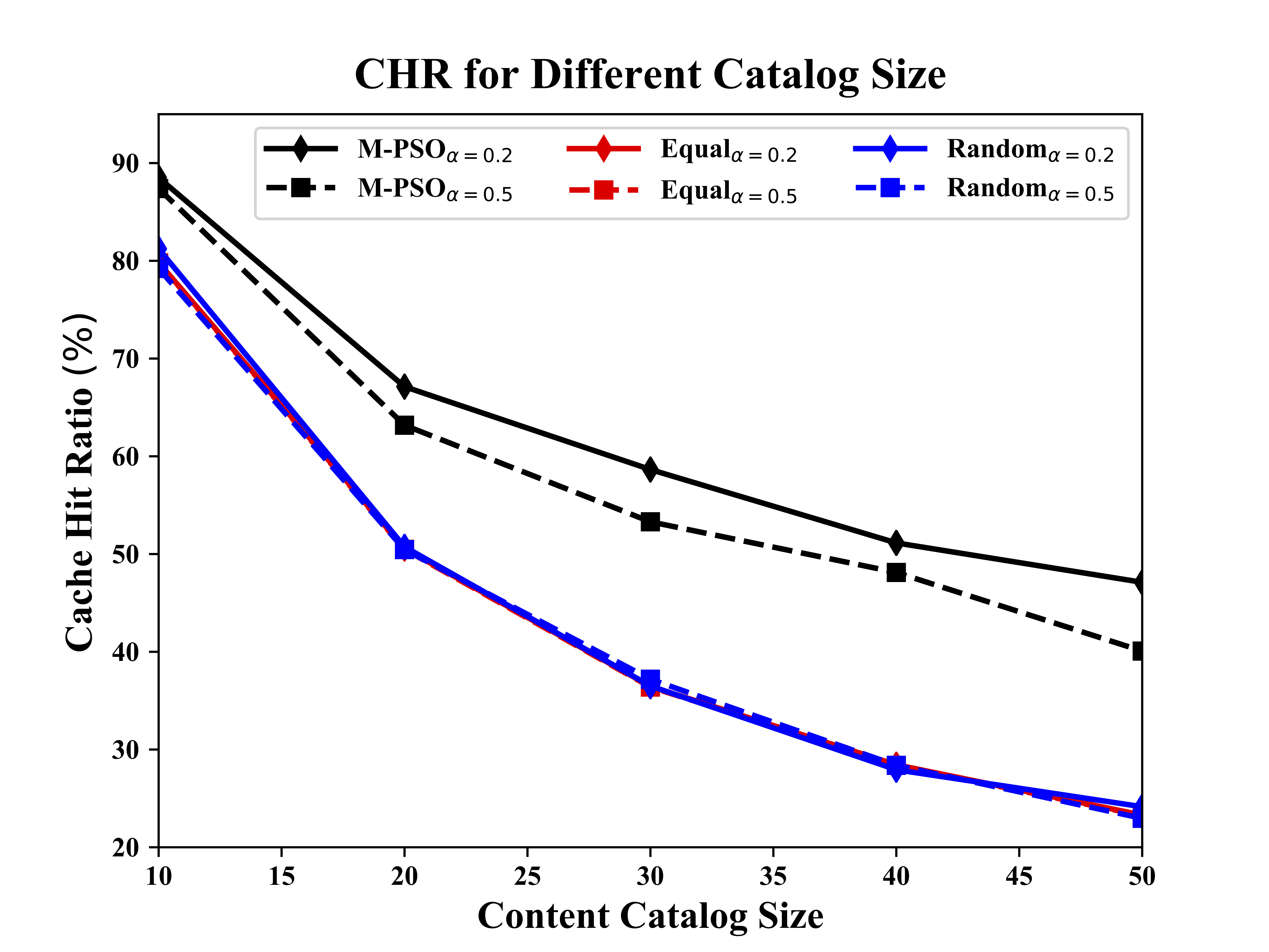}
	\caption{Impact of catalog size: CHR with $\mathcal{C}_d = 2$, $\mathcal{C}_b = 4$ and $\mathcal{C}_m = 8$}
	\label{Impact_of_Catalog}
\end{figure}

\begin{figure*}
	\centering
	\subfloat[CHR for different user cache storage sizes ] {\label{CHR_4_different_UE_CacheSizeWithCb4AndCm08} \includegraphics[width= 0.33 \textwidth] {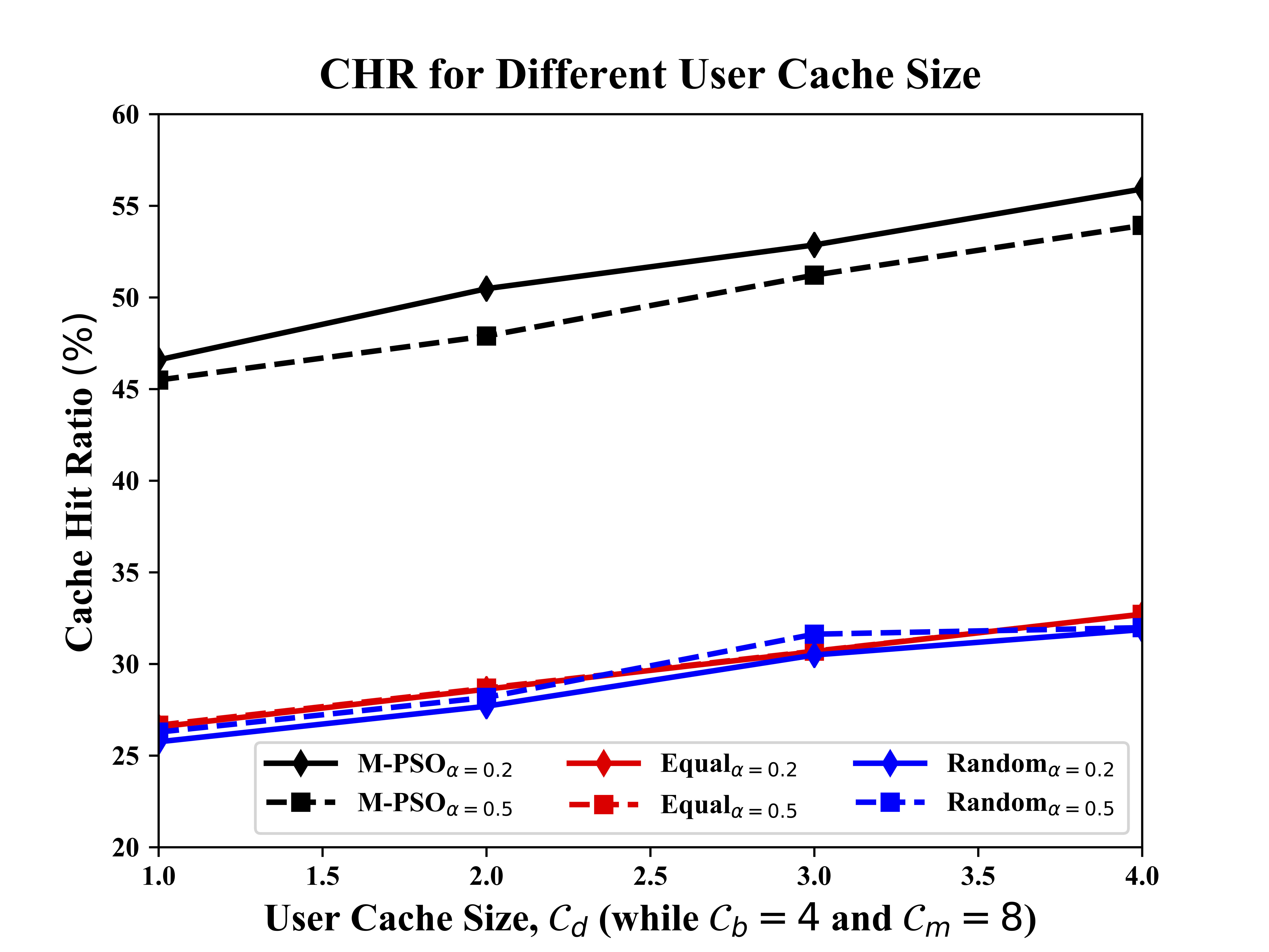}}
	\subfloat[CHR for different sBS cache storage sizes] {\label{CHR_4_different_sBS_CacheSizeWithCd2AndCm08} \includegraphics[width= 0.33 \textwidth] {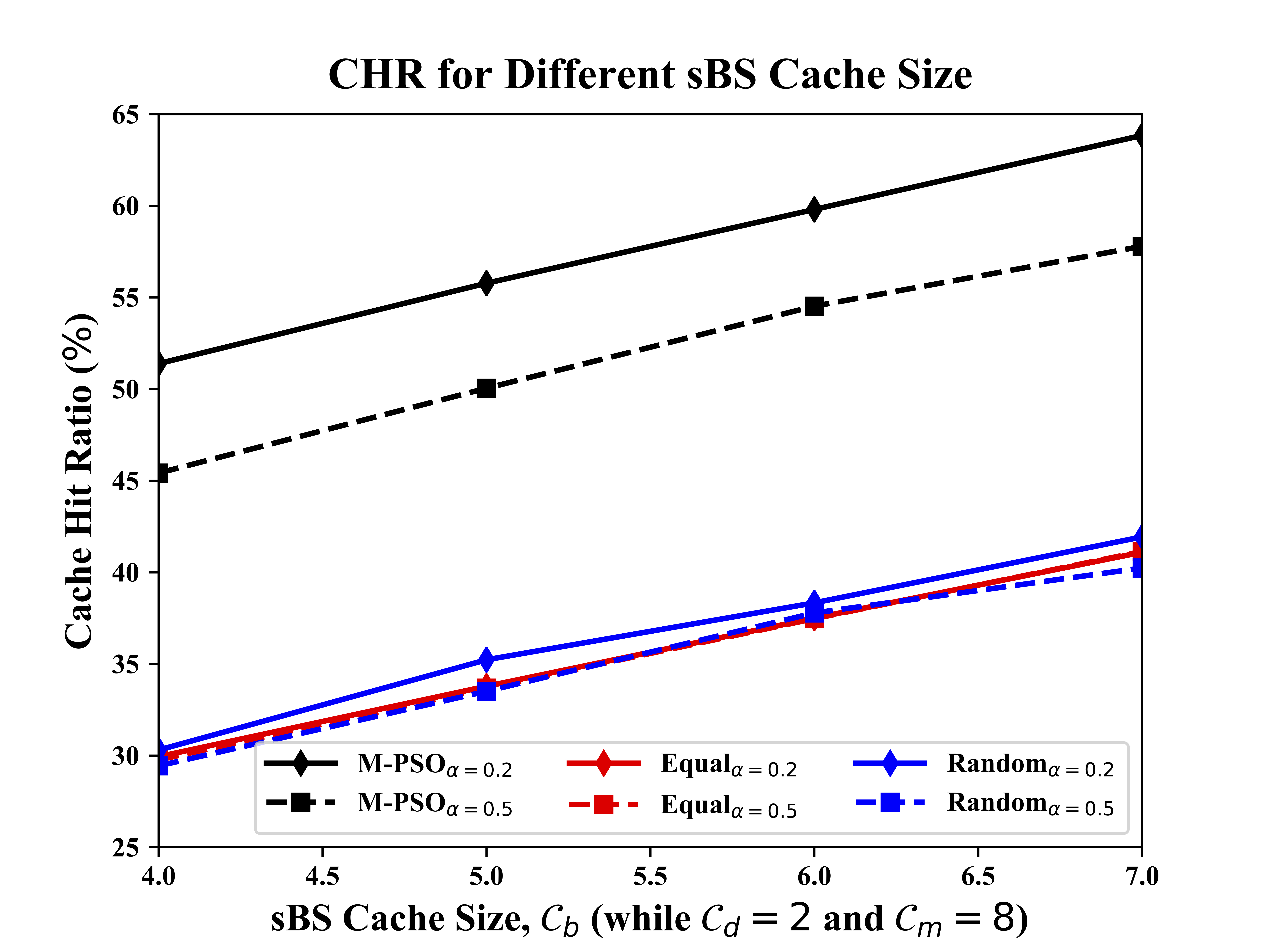}}
	\subfloat[CHR for different MBS cache storage sizes] {\label{CHR_4_different_MBS_CacheSizeWithCd2AndCb04} \includegraphics[width= 0.33 \textwidth] {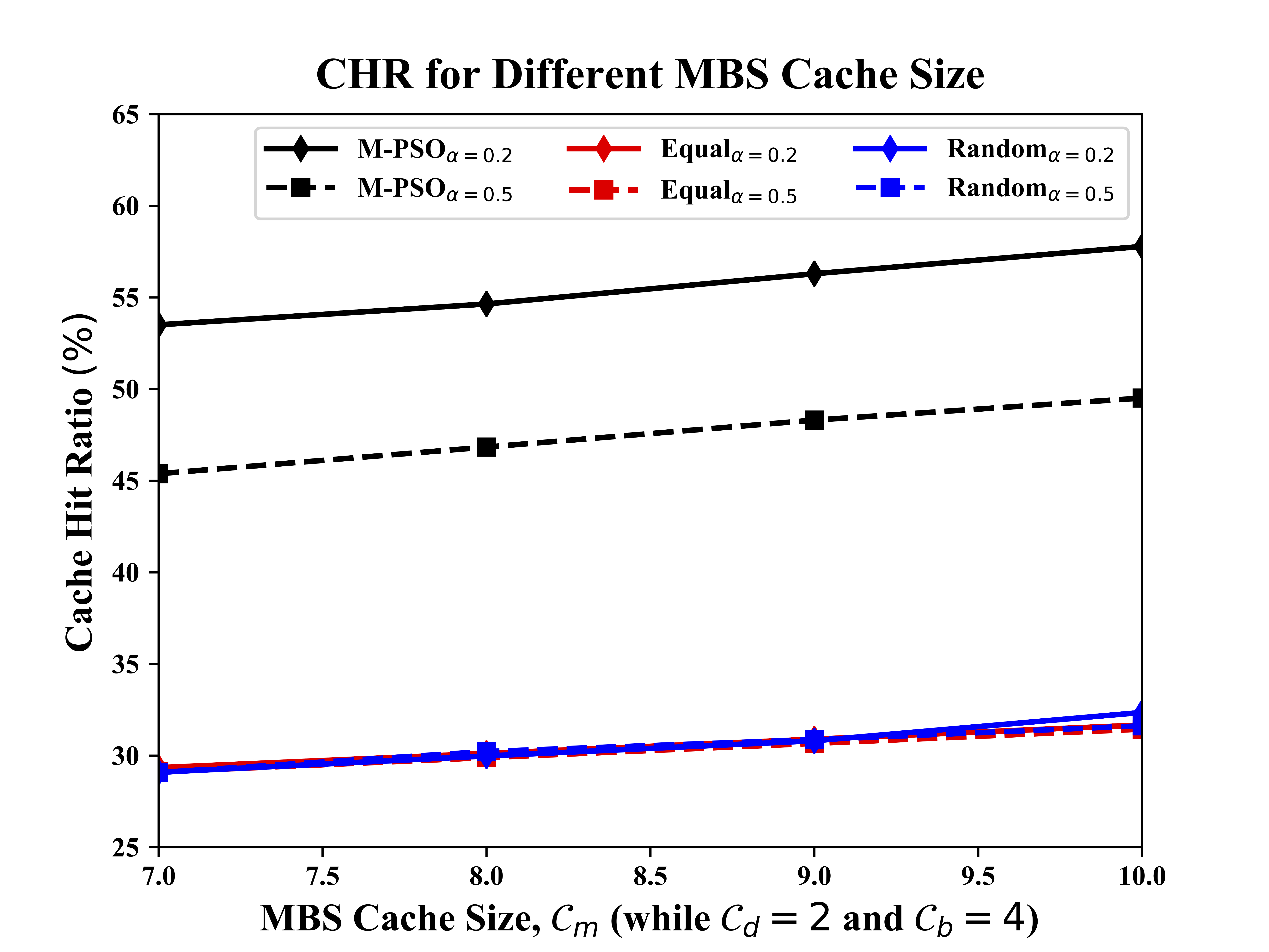}}
	\caption{Impact of cache size on CHR}
	\label{Impact_of_Cache_Size}
\end{figure*} 

\subsection{Performance Analysis}
We study the performance of our proposed M-PSO algorithm and make a fair comparison to the following benchmark caching schemes in this sub-section.

\textbf{Random Caching Scheme}:
In the random caching scheme, contents are stored randomly, while satisfying the constraints. 
  
\textbf{Equal Caching Scheme}: 
    In the equal caching scheme, each content is placed with the same probability. 

To show the effectiveness of our proposed algorithm, we only consider $100$ iterations. 
In Fig.~\ref{Comparision}, we demonstrate the results obtained from using our proposed algorithm, random caching scheme, and equal caching scheme. 
With only $100$ iterations, we achieve $\approx 24\%$ better performance than the benchmark caching schemes. 
Therefore, we claim that our proposed algorithm achieves better system performance than the other baseline caching schemes within a minimal number of iterations. 
In the following, we use our algorithm to evaluate the system performance in terms of different parameter setting.

\subsubsection{Impact of the Catalog Size}
Recall that if the requested content is delivered from one of the cache-enabled edge nodes, a cache hit occurs. 
Therefore, we aim to store as many to-be-requested contents as possible into the local edge nodes.
We consider the catalog size in the set of $[10, 20,30,40,50]$. 
Furthermore, the intensities are set as $\lambda_u = 10^{-4}$, $\lambda_b = 10^{-5}$ and $\lambda_m = 10^{-7}$. 
Also, the total number of iterations is chosen in the set of $100 \times [1, 10, 20, 40, 80]$ for the catalog size in [10, 20, 30, 40, 50], respectively. 
Note that if the catalog size increases, the number of possible combinations also increases. 
Therefore, whenever the content catalog increases, we slightly increase the total number of iterations. 
Also, if the total number of contents increases and we have only a limited number of cache-enabled nodes, the chance of storing the contents locally decreases, meaning that more content requests need to be served from the cloud. 
Therefore, the $\Sigma$ should decrease if the content catalog increases. 
Moreover, if the percentage of the requester nodes increases, the performance should degrade as we consider the heterogeneous preference of the users. 
Fig.~\ref{Impact_of_Catalog} also shows that if we increase the catalog size, $|\mathcal{F}|$ or the number of requesters ($\alpha$), then the $\Sigma$ decreases.

\subsubsection{Impact of the Storage Size}
We now investigate the impact of the cache sizes of the edge nodes on the system performance.
Remember that if cache size increases, more content can be stored at the cache-enabled nodes. 
Therefore, increasing the cache size of the users means that users store more contents at their local storage. 
As these storage sizes increases, the job of the proposed M-PSO algorithm is to determine the optimal caching placements. 
The simulation results, presented in Fig. \ref{Impact_of_Cache_Size}, validate that as the storage size increases more content is locally stored leading to an improved CHR.
Notice that increasing MBS cache size provides lesser CHR gain than increasing the cache size of the D2D users (or, the sBS).
This is because the total number of MBS are typically very lower than the available D2D (or, sBS) nodes.

\section{Conclusion}
\label{conclu}
Caching solution helps to achieve better system performances. 
However, the hard combinatorial decision-making problem of placing the contents at the local nodes is challenging. 
The grand problem is effectively solved with good accuracy by using the artificial intelligence  based technique. 
Considering heterogeneous content preferences in a real-world network platform, the proposed algorithm converges fast and achieves a much better performance than the existing benchmark caching schemes.

\section*{Acknowledgment}
The authors sincerely thank Shaju Shah for the critical and helpful discussions during this work.

\bibliography{Reference}
\bibliographystyle{IEEEtran}

\end{document}